\documentclass[%
 reprint,
 amsmath,amssymb,
 aps,
 prl,
 longbibliography,
 lengthcheck,%
]{revtex4-1}

\usepackage{graphicx}% Include figure files
\usepackage{dcolumn}% Align table columns on decimal point
\usepackage{bm}% bold math
\usepackage{hyperref}% add hypertext capabilities
\begin{document}

\preprint{APS/123-QED}

\title{Exact analytical soliton solutions in dipolar BEC
}% Force line breaks with \\

\author{P. A. Andreev}%
\email{andreevpa@physics.msu.ru}
\author{L. S. Kuz'menkov}
\email{lsk@phys.msu.ru}
 \affiliation{Physics Faculty, Moscow State
University, Moscow, Russian Federation.}%Lines break automatically or can be forced with \\

%\affiliation{%
% Department of Theoretical Physics, Physics Faculty, Moscow State
%University, Moscow, Russian Federation.}%


%\affiliation{
%\affiliation{
%\affiliation{%


\date{\today}% It is always \today, today,

\begin{abstract}
The bright, dark and grey solitons are well-known soliton solutions of the Gross-Pitaevskii equation for the attractive and repulsive BEC. We consider solitons in the dipolar BEC of the fully polarized particles, speaking of the dipolar BEC we mean both the magnetized BEC and the electrically polarized BEC. We show that these two types of the dipolar BEC reveal different behavior of the collective excitations. This is related to the fact that the electric and the magnetic fields satisfy to the different pairs of the Maxwell equation set. Thus we consider them independently. We obtain exact analytical solutions for the bright, dark, and grey solitons in the magnetized (electrically polarized) BEC when they propagate parallel and perpendicular to an external magnetic (electric) field. Comparison of spectrum of the linear collective excitations for the two kinds of the dipolar BEC is presented as well.
\end{abstract}

%\pacs{03.75.Kk  03.75.Hh}% PACS, the Physics and Astronomy
%Use showkeys class option if keyword
\maketitle

%\tableofcontents

\section{Introduction}

Solitons as fundamental nonlinear excitations in dipolar BEC have been actively studied \cite{Nath PRL 08}-\cite{Rojas PRA 11}. These studies are based on using of the generalization of the Gross-Pitaevskii (GP) equation for dipolar BEC \cite{Yi PRA 00}-\cite{Carr NJP 09}. Electrically polarized BEC was also considered as a dielectric medium \cite{Wilson NJP 12}, an electric field caused by polarization was explicitly introduced there. It has been shown that at consideration of the GP equation for dipolar BEC whole expression for potential energy of dipole-dipole interaction have to be used explicitly \cite{Andreev 2013 non-int GP}. It allows to introduce electric field caused by dipoles, which satisfy to the Maxwell equations. In this case the GP equation for the dipolar BEC can be presented in an nonintegral form, and it is coupled with the pair of the Maxwell equations. This was developed for the electrically polarized BEC \cite{Andreev 2013 non-int GP}. In this paper we will show that analogous representation of the GP equation for the magnetized BEC can be done. We present an explicit expression for the potential energy of magnetic dipole interaction, or in other words the energy of the spin-spin interaction, which leads to corresponding Maxwell equations. We derive our model from the first principles (the many-particle Schrodinger equation) using method of the quantum hydrodynamics (QHD) \cite{MaksimovTMP 2001}-\cite{Shukla RMP 11}, which has been used for studying of the dipolar BEC \cite{Andreev arxiv Pol}-\cite{Andreev TransDipBEC12}. However, dynamics of dipole direction had been included in previous research, so it was more complicated model hiding some features of the model, which are necessary to understand.

In this paper we develop reduced form of the QHD equations for fully polarized BECs. Nevertheless, even reduced model reveals different evolution of electrically polarized and magnetized BECs. Moreover, we show that nonlinear Schrodinger equations obtained in our model from the reduced QHD equations differ from well-known generalization of the GP equation has been used for the dipolar BEC description (see, for instance \cite{Lahaye RPP 09}, \cite{Baranov CR 12}). This leads to different properties of collective excitations for the two kinds of dipolar BEC, and to difference of they properties from obtained using the integral GP equation, which are  described in review papers \cite{Lahaye RPP 09}, \cite{Baranov CR 12}).

We study dispersion of collective excitations for both kinds of the fully polarized dipolar BECs, as for the magnetized (spinor) BEC and for the electrically polarized BEC. We discuss differences of dispersion dependencies and reasons laying behind these differences. The reduced QHD leads to a nonintegral GP equations for the dipolar BECs, just as the reduced QHD equations can be presented in nonintegral form instead of integral GP (see Ref. \cite{Lahaye RPP 09} formula (4.3)) and integral Euler (see Ref. \cite{Lahaye RPP 09} formula (4.9)) equations. We show that these nonintegral equations are powerful method for finding analytical solutions for solitons in dipolar BECs. We consider twelve cases allowing to find analytical soliton solutions. We obtain solutions for the bright, dark, and grey solitons propagating parallel (perpendicular) to external field as for the magnetized and for the electrically polarized BECs.

%\begin{equation}\label{SDBEC} \end{equation}

The paper is organized as follows. In Sec. II is dedicated to description of our model. We briefly show the scheme of derivation of our model from the first principles of microscopic theory. We describe equations of the quantum hydrodynamic for spinning particles (spinor BEC) and for particles having electric dipole moment being in the BEC state. We describe differences between properties of the two kind of BECs. We show how these equations reduce for the fully polarized BECs. We present and prove correct form of the potential energy for interaction of the electric dipoles, and for the spin-spin interaction, which we use for derivation of the QHD equations. In Sec. III we calculate and discuss dispersion of the linear collective excitations in the two kind of the dipolar BEC. In Sec. IV we make a brief review of solitons in an uniform unpolarized BEC, we present solutions for the bright, dark, and grey solitons, which we use in the following sections. In Sec. V we describe the solitons in the magnetized BEC, we consider propagation of the solitons parallel and perpendicular to an external magnetic field. In Sec. VI we study the solitons in the electrically polarized BEC considering propagation parallel and perpendicular to an external electric field. In Sec. VII we recapture results of our paper.

\section{Model}

\subsection{Quantum hydrodynamic equations for magnetized BEC}

The Schrodinger
equation describing dynamic of $N$ spinning particles being involved in the long-range spin-spin interaction and in the short-range interaction, \emph{and} interacting with an external magnetic field is
$$\imath\hbar\partial_{t}\psi_{s}(R,t)=\Biggl(\sum_{i}\biggl(\frac{\textbf{p}^{2}_{i}}{2m_{i}}-\mu\hat{s}^{\alpha}_{i}B^{\alpha}_{i(ext)}\biggr) $$
\begin{equation}\label{SDBEC magn BEC spin gam}+\frac{1}{2}\sum_{i,j\neq i}\biggl(U_{ij}-\mu^{2}G^{\alpha\beta}_{ij}\hat{s}^{\alpha}_{i}\hat{s}^{\beta}_{j}\biggr)\Biggr)\psi_{s}(R,t),\end{equation}
where $\mu$ is the magnetic moment of particle,
$\textbf{p}_{i}=-\imath\hbar\partial_{i}$ is the operator of
momentum, $\partial_{i}$ is the space derivative on coordinate of $i$th particle, $\textbf{B}_{i(ext)}$ is the external magnetic field acting on $i$th particle, $U_{ij}$ present the short-range interaction (SRI), the Green
function of the spin-spin interaction (SSI) has form
\begin{equation}\label{SDBEC magn BEC SSI Green function}G^{\alpha\beta}_{ij}=\partial^{\alpha}_{i}\partial^{\beta}_{i}(1/r_{ij})+4\pi\delta_{\alpha\beta}\delta(\textbf{r}_{ij}),\end{equation}
for spin matrixes $\hat{s}^{\alpha}_{i}$ the commutation relations
are
$$[\hat{s}^{\alpha}_{i},\hat{s}^{\beta}_{j}]=\imath\delta_{ij}\varepsilon^{\alpha\beta\gamma}\hat{s}^{\gamma}_{i}.$$
The first term in the right-hand side of the Schrodinger equation (\ref{SDBEC magn BEC spin gam}) presents the kinetic energy of particles, the second term is the potential energy of magnetic moments in an external magnetic field $\textbf{B}_{i, ext}$, subindex $i$ shows that field $\textbf{B}_{i, ext}$ acts on $i$-th particle meaning that in the nonuniform magnetic field different field acts on different particles. The third and fourth terms describe the interparticle interaction, the short-range and the spin-spin interactions, correspondingly.

We consider spin-1 Bose particles, thus we present the explicit
form of the spin matrixes $\hat{s}^{\alpha}_{i}$ for particles
with spin equal to 1:
$$\begin{array}{ccc} \hat{s}_{x}=\frac{1}{\sqrt{2}}\left(\begin{array}{ccc}0&
1&
0\\
1& 0&
1\\
0& 1&
0\\
\end{array}\right),&
\hat{s}_{y}=\frac{1}{\sqrt{2}}\left(\begin{array}{ccc}0& -\imath &
0\\
\imath & 0&
-\imath \\
0& \imath &
0\\
\end{array}\right),&
\end{array}$$
$$\hat{s}_{z}=\left(\begin{array}{ccc}1&
0&
0\\
0& 0&
0\\
0& 0&
-1\\
\end{array}\right).$$

Starting from the many-particle Schrodinger equation (\ref{SDBEC magn BEC spin gam}) and using the QHD method we derive a set of equations for spinning particles in the BEC state. The set of equations consists of three equations, they are the continuity equation, the Euler equation giving momentum balance, and the magnetic moment (spin) evolution equation, which is a generalization of the Bloch equation, the last one describes dynamic of single spin in an external magnetic field.

The first and basic step at getting of the QHD equations from the many-particle wave function is the definition of the particle concentration $n(\textbf{r},t)$ as the quantum mechanical average of the concentration operator in the coordinate presentation of the quantum mechanics $\hat{n}=\sum_{i=1}^{N}\delta(\textbf{r}-\widehat{\textbf{r}}_{i})$, so we have
\begin{equation}\label{SDBEC concentration definition} n(\textbf{r},t)=\int \psi^{*}(R,t)\hat{n}\psi(R,t) dR, \end{equation}
where $dR=\prod_{n=1}^{N}d\textbf{r}_{n}$ is a infinitesimally small volume in 3N dimensional configurational space. Next, differentiating the particles concentration (\ref{SDBEC concentration definition}) with respect to time we get the continuity equation and an explicit form of the particle current $\textbf{j}=n\textbf{v}$, which is not presented here but can be found in papers dedicated to details of equation derivation, see for instance \cite{MaksimovTMP 2001}, \cite{Andreev PRA08}, \cite{Andreev ch in B (A)}, \cite{Andreev IJMP B 13}, \cite{Andreev PRB 11}. Explicit form of other quantum hydrodynamical variables is also presented in mentioned papers. Having explicit form of the particles current $\textbf{j}$ and magnetization $\textbf{M}$ we can derive corresponding equations of evolution, which are presented below. The reader, who wants to know some details of the derivation of the QHD equation, we suggest to take following papers. The QHD equations for spinning particles were firstly derived in Ref. \cite{MaksimovTMP 2001} for the charged spinning particles. General technic of the derivation of the QHD equations for the spinless particles can be found in Ref.s \cite{Andreev PRA08}, \cite{Andreev ch in B (A)}, and \cite{Andreev PRB 11}. In Ref.s \cite{Andreev PRA08} and \cite{Andreev ch in B (A)} attention was stressed on the short-range interaction of neutral particles in the BEC state.

\subsubsection{General form of hydrodynamic equations and introduction of magnetic field caused by magnetic moments}

The continuity equation
\begin{equation}\label{SDBEC cont eq}\partial_{t}n+\partial^{\alpha}(nv^{\alpha})=0\end{equation}
shows conservation of the particle number and describes time evolution of the particle concentration.

The momentum balance equation
$$mn(\partial_{t}+\textbf{v}\nabla)v^{\alpha}-\frac{\hbar^{2}}{4m}\partial^{\alpha}\triangle
n$$
$$+\frac{\hbar^{2}}{4m}\partial^{\beta}\Biggl(\frac{\partial^{\alpha}n\cdot\partial^{\beta}n}{n}\Biggr)-\frac{\hbar^{2}}{4m\mu^{2}}\partial^{\beta}\Biggl(M^{\gamma}\partial^{\alpha}\partial^{\beta}\biggl(\frac{M^{\gamma}}{n}\biggr)\Biggr)$$
\begin{equation}\label{SDBEC bal imp eq full}=-gn\nabla n+M^{\beta}\partial^{\alpha}\biggl(B^{\beta}_{ext}(\textbf{r},t)+\int
d\textbf{r}'G^{\beta\gamma}(\textbf{r},\textbf{r}')M^{\gamma}(\textbf{r}',t)\biggr)
\end{equation}
appears in the rather complicated form. The first group of terms $mn(\partial_{t}+\textbf{v}\nabla)\textbf{v}=mn\frac{d\textbf{v}}{dt}$ is the convective time derivative of the velocity field $\textbf{v}$. Next three terms are proportional to $\hbar^{2}$ and present the quantum Bohm potential for spinning particles, the first two of them are the quantum Bohm potential for spinless particles corresponding to the GP equation for scalar macroscopic wave function. The last of them arises due to the spin. If spin of all particles are parallel to each other and to an external magnetic field, we can rewrite the magnetization $\textbf{M}$ as $\textbf{M}=\mu n \textbf{l}$, where $\textbf{l}$ is a unit vector in the direction of magnetic moment, which is parallel to direction of the external magnetic field, $\mu$ is the magnetic moment of particle, $n$ is the particle concentration. In described approximation evolution of the magnetization reduces to the particle concentration evolution. So, discussed term reduces to zero $\partial^{\beta}\Biggl(M^{\gamma}\partial^{\alpha}\partial^{\beta}\biggl(\frac{M^{\gamma}}{n}\biggr)\Biggr)$$=\mu^{2}\partial^{\beta}(n\textbf{l}\partial^{\alpha}\partial^{\beta}\textbf{l})=0$, since $\textbf{l}$ is a constant. The right-hand side of equation (\ref{SDBEC bal imp eq full}) describes the force density leading to the particle evolution. The first term describes the short-range interaction in the first order by the interaction radius \cite{Andreev PRA08}, \cite{Andreev ch in B (A)}, \cite{Andreev IJMP B 13}, \cite{L.P.Pitaevskii RMP 99}. The next group of terms presents interaction of the magnetic moment with the external magnetic field and interaction between the magnetic moments (the spin-spin interaction), correspondingly.

Magnetization evolution is described by the following equation
$$\partial_{t}M^{\alpha}+\partial^{\beta}\Biggl(M^{\alpha}v^{\beta}-\frac{\hbar}{2m\mu}\varepsilon^{\alpha\mu\nu}M^{\mu}\partial^{\beta}\biggl(\frac{M^{\nu}}{n}\biggr)\Biggr)$$
\begin{equation}\label{SDBEC magn evol eq}=\frac{\mu}{\hbar}\varepsilon^{\alpha\beta\gamma}M^{\beta}\biggl(B^{\gamma}_{ext}(\textbf{r},t)+\int
d\textbf{r}'G^{\gamma\delta}(\textbf{r},\textbf{r}')M^{\delta}(\textbf{r}',t)\biggr).\end{equation}
The second term $\partial^{\beta}(M^{\alpha}v^{\beta})$ is an analog of convective part of the time derivative of the magnetization, which is also convective part of the magnetization flow. The third term is the quantum part of the magnetization flow, which is an analog of the quantum Bohm potential in the momentum balance equation (\ref{SDBEC bal imp eq full}). Let us discuss now the right-hand side of magnetic moment evolution equation (\ref{SDBEC magn evol eq}). The terms in the right-hand side give contribution of interaction in time evolution of magnetization. The first term presents interaction of magnetic moments with the external magnetic field, and the last term describes spin-spin interaction. We admit that the short-range interaction, when interaction between particles with different spin projection is described by the same potential, gives no contribution in this equation.

Set of the QHD equations (\ref{SDBEC cont eq})-(\ref{SDBEC magn evol eq}) is a closed set of equations. However these equations contain integral terms describing the interparticle interaction. We can see, from equations (\ref{SDBEC bal imp eq full}) and (\ref{SDBEC magn evol eq}), that  the integral terms are written along with the external magnetic field, so we can expect that these terms describe an internal magnetic field caused by the magnetic moments. Thus, we can explicitly write the internal magnetic field
\begin{equation}\label{SDBEC magnetic field} B^{\alpha}(\textbf{r},t)=\int d\textbf{r}' G^{\alpha\beta}(\mid\textbf{r}-\textbf{r}'\mid)M^{\beta}(\textbf{r}',t),\end{equation}
and find that this field satisfies to the Maxwell equations
\begin{equation}\label{SDBEC magn BEC Maxwell curl}\nabla\times \textbf{B}(\textbf{r},t)=4\pi \nabla\times \textbf{M}(\textbf{r},t),\end{equation}
and
\begin{equation}\label{SDBEC magn BEC Maxwell div}\nabla \textbf{B}(\textbf{r},t)=0.\end{equation}
Consequently, the first term in the right-hand side of equation (\ref{SDBEC bal imp eq full}) can be written as $M^{\beta}\nabla B^{\beta}$, where $\textbf{B}$ is the sum of the external and internal magnetic fields. Analogously, the right-hand side of the magnetic moment evolution equation appears as $(\mu/\hbar)\varepsilon^{\alpha\beta\gamma}M^{\beta}B^{\gamma}$. As a result we have a set of nonintegral equation explicitly including the magnetic field caused by the magnetic moments coupled with the equations of field (\ref{SDBEC magn BEC Maxwell curl}) and (\ref{SDBEC magn BEC Maxwell div}).

\subsubsection{Reduced set of QHD equations for the fully magnetized BEC}

For the fully magnetized BEC, with no evolution of the magnetic moment direction, we have no change in the continuity equation (\ref{SDBEC cont eq}), the magnetic moment evolution equation disappears in this limit, and we write down a reduced Euler equation
$$mn(\partial_{t}+\textbf{v}\nabla)v^{\alpha}-\frac{\hbar^{2}}{4m}\partial^{\alpha}\triangle
n$$
\begin{equation}\label{SDBEC bal imp eq fully magn}+\frac{\hbar^{2}}{4m}\partial^{\beta}\Biggl(\frac{\partial^{\alpha}n\cdot\partial^{\beta}n}{n}\Biggr)=-gn\nabla^{\alpha}n+\mu n\partial^{\alpha}\textbf{l}\textbf{B},\end{equation}
where we have used that $M^{\beta}\nabla B^{\beta}=\mu n l^{\beta}\nabla B^{\beta}$$=\mu n\nabla (\textbf{l}\textbf{B})$.

Moreover, for the fully magnetized BEC equations of field represent as
\begin{equation}\label{SDBEC magn div short}\nabla\textbf{B}=0,\end{equation}
and
\begin{equation}\label{SDBEC magn curl short}\nabla\times\textbf{B}=4\pi\mu\nabla n\times \textbf{l}.\end{equation}

We can also put down corresponding NLSE, which can be directly derived from couple of the continuity and the Euler equations (\ref{SDBEC cont eq}) and (\ref{SDBEC bal imp eq fully magn}) (see Ref. \cite{Andreev PRA08}),
\begin{equation}\label{SDBEC nlse int polariz some frame non Int}\imath\hbar\partial_{t}\Phi(\textbf{r},t)=\Biggl(-\frac{\hbar^{2}}{2m}\triangle+g\mid\Phi(\textbf{r},t)\mid^{2}-\mu\textbf{l}\textbf{B}(\textbf{r},t)\Biggr)\Phi(\textbf{r},t),\end{equation}
where the last term $-\mu\textbf{l}\textbf{B}$ describes the spin-spin interaction and interaction of the magnetic moments with the external magnetic field. This term appears instead of the integral term in the generalization of the GP equation for dipolar BEC (see, for instance, Ref.s \cite{Lahaye RPP 09} and \cite{Baranov CR 12}). An equation analogous to (\ref{SDBEC nlse int polariz some frame non Int}) was obtained for the electrically polarized BEC \cite{Andreev 2013 non-int GP}.

\subsubsection{Potential energy of spin-spin interaction}

Potential energy of the spin-spin interaction was directly derived from the quantum electrodynamical scattering amplitude of two electrons \cite{spin-spin interaction}. This potential energy consists of two parts. The first one is well-known appearing as a fraction \begin{equation}\label{SDBEC d-d Ham fraction only}U_{dd}=\frac{\textbf{d}^{2}-3(\textbf{d}\textbf{r})^{2}/r^{2}}{r^{3}},\end{equation}
and the second term is proportional to the Dirac delta function. However, coefficient at the delta function for the spin-spin interaction differs from one for the spin-spin interaction. Since, the magnetic field caused by magnetic dipole should satisfy to the Maxwell equations $div \textbf{B}=0$ and $curl \textbf{B}=4\pi curl \textbf{M}$, which differ from the Maxwell equations for the electric field, the last of these equations is written in the quasi static approximation and in absence of electric currents. Finally, we write down different forms of the Hamiltonian of spin-spin interaction
$$H_{\mu\mu}=\Biggl(\frac{\delta^{\alpha\beta}-3r^{\alpha}r^{\beta}/r^{2}}{r^{3}}-\frac{8\pi}{3}\delta^{\alpha\beta}\delta(\textbf{r})\Biggr)\mu^{\alpha}_{i}\mu^{\beta}_{j}.$$
This is explicit form of
\begin{equation}\label{SDBEC pot of mm int} H_{\mu\mu}=-\Biggl(4\pi\delta^{\alpha\beta}\delta(\textbf{r}_{ij})+\nabla^{\alpha}_{i}\nabla^{\beta}_{i}\frac{1}{r_{ij}}\Biggr)\mu^{\alpha}_{i}\mu^{\beta}_{j},\end{equation}
which we can rewrite as
\begin{equation}\label{SDBEC}H_{\mu\mu}=-\mu^{\alpha}_{i}\mu^{\beta}_{j}\Biggl(\partial^{\alpha}_{i}\partial^{\beta}_{i}-\delta^{\alpha\beta}\Delta_{i}\Biggr)\frac{1}{r_{ij}}.\end{equation}
Explicit form of the Hamiltonian is obtained using following identities
\begin{equation}\label{SDBEC togdestvo}-\partial^{\alpha}\partial^{\beta}\frac{1}{r}= \frac{\delta^{\alpha\beta}-3r^{\alpha}r^{\beta}/r^{2}}{r^{3}}+\frac{4\pi}{3}\delta^{\alpha\beta}\delta(\textbf{r}),\end{equation}
and
\begin{equation}\label{SDBEC togdestvo scalar}\triangle\frac{1}{r}= -4\pi\delta(\textbf{r}).\end{equation}

\subsection{Quantum hydrodynamic equations for electrically polarized BEC}

Let us start with discussion of potential energy of electric dipole interactions. In previous subsection we have met the Hamiltonian of spin-spin interaction, which differs from usually using form of the Hamiltonian of dipole-dipole interaction (\ref{SDBEC d-d Ham fraction only}). Here we are going to show that the potential energy of electric dipole interaction differs from (\ref{SDBEC d-d Ham fraction only}) and (\ref{SDBEC pot of mm int}). Potential of the electric field caused by an electric dipole appears as $\varphi=-(\textbf{d}\nabla)(1/r)$ \cite{Landau 2}, consequently the electric field has form of $\textbf{E}=-\nabla\varphi$$=(\textbf{d}\nabla)\nabla(1/r)$ or it can be rewritten in the tensor form $E^{\alpha}=d^{\beta}\nabla^{\beta}\nabla^{\alpha}(1/r)$. Energy of interaction of two dipoles comes as $U_{dd}=-\textbf{d}_{j}\textbf{E}_{ji}$, where $\textbf{d}_{j}$ is the electric dipole moment of the second dipole, and $\textbf{E}_{ji}$ is the electric field caused by the first dipole acting on the second dipole. Explicit form of the dipole-dipole interaction energy is $U_{dd}=-d^{\alpha}_{j}d^{\beta}_{i}\nabla^{\beta}_{i}\nabla^{\alpha}_{i}(1/r_{ij})$. It can be rewritten via notion of the Green function $G^{\alpha\beta}$ of dipole-dipole interaction $U_{dd}=-d^{\alpha}_{j}d^{\beta}_{i}G^{\alpha\beta}_{ij}$, where $G^{\alpha\beta}_{ij}=\partial^{\alpha}_{i}\partial^{\beta}_{i}\frac{1}{r_{ij}}$.

For consideration of the neutral particles having electric dipole moment $d_{i}^{\alpha}$ in the BEC state we derive a set of QHD equations. We starting from the many-particle Schrodinger equation
$$\imath\hbar\partial_{t}\psi(R,t)=\hat{H}\psi(R,t),$$
where $R=(\textbf{r}_{1}, ..., \textbf{r}_{N})$ is the set of coordinate of $N$ particles, with the Hamiltonian
$$\hat{H}=\sum_{i}\Biggl(\frac{1}{2m_{i}}\textbf{p}_{i}^{2}-d_{i}^{\alpha}E_{i,ext}^{\alpha}\Biggr)$$
\begin{equation}\label{SDBEC Hamiltonian}+\frac{1}{2}\sum_{i,j\neq i}\Biggl(U_{ij}-d_{i}^{\alpha}d_{j}^{\beta}G_{ij}^{\alpha\beta}\Biggr),\end{equation}
where $E_{i,ext}^{\alpha}$ is the
electric field, $m_{i}$ is mass of particles, $\hbar$ is the Planck constant, and
$G_{ij}^{\alpha\beta}=\partial_{i}^{\alpha}\partial_{i}^{\beta}1/r_{ij}$
is the Green functions of dipole-dipole
interaction discussed above, $U_{ij}=U(|\textbf{r}_{i}-\textbf{r}_{j}|)$ is the potential of the short-range interaction. Let us admit that the dipole-dipole interaction leads to evolution of both the positions of particles and the directions of electric dipole moments, but in this paper we neglect the evolution of dipole direction. Consequently we will simplify the Euler equation and get a closed set of the QHD equations.

Analogously for spinning particles we find the continuity
\begin{equation}\label{SDBEC cont eq for ED}
\partial_{t}n+\nabla(n\textbf{v})=0,\end{equation}
and the Euler
$$mn(\partial_{t}+\textbf{v}\nabla)v^{\alpha}-\frac{\hbar^{2}}{4m}\partial^{\alpha}\triangle n$$
$$+\frac{\hbar^{2}}{4m}\partial^{\beta}\biggl(\frac{1}{n}(\partial^{\alpha}n)(\partial^{\beta}n)\biggr)=-gn\partial^{\alpha}n
$$
\begin{equation}\label{SDBEC bal imp eq ED general} +P^{\beta}\partial^{\alpha}E_{ext}^{\beta}+P^{\beta}\partial^{\alpha}\int d\textbf{r}'G^{\beta\gamma}(\textbf{r},\textbf{r}')P^{\gamma}(\textbf{r}',t)\end{equation}
equations, where $n$ is the particle concentration, $\textbf{v}$ is the velocity field, $P^{\beta}$ is the polarization of the medium, or the density of the electric dipole moment. The last term in the Euler equation (\ref{SDBEC bal imp eq ED general}) contains the integral presenting internal electric field caused by the polarization of medium, this electric field has form
$$E^{\alpha}_{int}(\textbf{r},t)=\int d\textbf{r}' G^{\alpha\beta}(\textbf{r},\textbf{r}')P^{\beta}(\textbf{r}',t).$$
We can obtain that this electric field satisfies to the Maxwell equations
\begin{equation}\label{SDBEC field good div}\nabla\textbf{E}(\textbf{r},t)=-4\pi \nabla\textbf{P}(\textbf{r},t),\end{equation}
and
\begin{equation}\label{SDBEC field good curl}\nabla\times\textbf{E}(\textbf{r},t)=0.\end{equation}

For the fully polarized BEC of particles having electric dipole moment the Euler equation (\ref{SDBEC bal imp eq ED general}) reduces to
$$mn(\partial_{t}+\textbf{v}\nabla)v^{\alpha}-\frac{\hbar^{2}}{4m}\partial^{\alpha}\triangle n
+\frac{\hbar^{2}}{4m}\partial^{\beta}\biggl(\frac{1}{n}(\partial^{\alpha}n)(\partial^{\beta}n)\biggr)$$
\begin{equation}\label{SDBEC bal imp eq short from GP}=-gn\partial^{\alpha}n+
nd\partial^{\alpha}(\textbf{l}\textbf{E}),\end{equation}
where $\textbf{l}$ is the unit vector in direction of polarization formed by external field $\textbf{P}=dn\textbf{l}$. Along with the Euler equation one of the Maxwell equations reduces too, so we have
\begin{equation}\label{SDBEC field good div simplified}\nabla\textbf{E}(\textbf{r},t)=-4\pi d(\textbf{l}\nabla) n(\textbf{r},t),\end{equation}
instead of equation (\ref{SDBEC field good div}). Equation (\ref{SDBEC field good curl}) has no change.

The generalization of the GP equation, for particles having electric dipole moment, corresponding to equation (\ref{SDBEC bal imp eq short from GP}) appears as
\begin{equation}\label{SDBEC nlse int polariz some frame non Int ED}\imath\hbar\partial_{t}\Phi(\textbf{r},t)=\Biggl(-\frac{\hbar^{2}}{2m}\triangle+g\mid\Phi(\textbf{r},t)\mid^{2}-d\textbf{l}\textbf{E}(\textbf{r},t)\Biggr)\Phi(\textbf{r},t),\end{equation}
where the last term is proportional to $-d\textbf{l}\textbf{E}$, it describes both the interaction among dipoles and the interaction of dipoles with the external electric field, this term appears instead of the integral term (see, for instance, Ref. \cite{Lahaye RPP 09} formula (4.3), \cite{Baranov CR 12} formula (15), and \cite{Carr NJP 09} formula (7)). We see that equation (\ref{SDBEC nlse int polariz some frame non Int ED}) has the same structure as equation (\ref{SDBEC nlse int polariz some frame non Int}) for spinning particles, but in equation (\ref{SDBEC nlse int polariz some frame non Int}) we have the magnetic moment $\mu$ instead of the electric dipole moment $d$, and the magnetic field $\textbf{B}$ instead of the electric field $\textbf{E}$. Difference in evolution of the two kinds of BEC reveals in the Maxwell equations for the electric $\textbf{E}$ and magnetic $\textbf{B}$ fields (\ref{SDBEC magn div short}), (\ref{SDBEC magn curl short}), (\ref{SDBEC field good curl}), and (\ref{SDBEC field good div simplified}). Equation (\ref{SDBEC nlse int polariz some frame non Int ED}) was obtained in Ref. \cite{Andreev 2013 non-int GP}.

\section{Linear excitations}

We find that the spectrum of collective excitations in the fully magnetized and in the fully electrically polarized are different. For the electrically polarized BEC we have
\begin{equation}\label{SDBEC disp for ED}\omega^{2}=\frac{\hbar^{2}k^{4}}{4m^{2}}+\frac{gn_{0}k^{2}}{m}+\frac{4\pi n_{0}d^{2}k^{2}\cos^{2}\theta}{m},\end{equation}
where $\cos\theta=k_{z}/k$, for details of calculation see Ref. \cite{Andreev 2013 non-int GP},
and for the magnetized BEC spectrum appears as
\begin{equation}\label{SDBEC disp for MD}\omega^{2}=\frac{\hbar^{2}k^{4}}{4m^{2}}+\frac{gn_{0}k^{2}}{m}-\frac{4\pi n_{0}\mu^{2}k^{2}\sin^{2}\theta}{m}.\end{equation}
Details of the calculations for the collective excitation spectrum of the magnetized BEC are presented in Appendix.

Formulas (\ref{SDBEC disp for ED}) and (\ref{SDBEC disp for MD}) are generalization of the Bogoliubov spectrum for the fully polarized dipolar BECs, for the BEC of particles having electric dipole moment and for the BEC of spinning particles, correspondingly. The first term in the right-hand side of equations (\ref{SDBEC disp for ED}) and (\ref{SDBEC disp for MD}) describes dispersion of the free quantum particle (dispersion of the de Broglie wave), it appears from the quantum Bohm potential in the Euler equation (see equation (\ref{SDBEC bal imp eq fully magn}) for the magnetized BEC, and equation (\ref{SDBEC bal imp eq short from GP}) for the electrically polarized BEC). The second term describes the short-range interaction and contains the constant of short-range interaction $g$, which does not contain contribution of the dipole-dipole interaction. It depends neither on the electric dipole moment $d$ no on the magnetic moment $\mu$. The last term in formula (\ref{SDBEC disp for ED}) (in formula (\ref{SDBEC disp for MD})) appears due to interaction of the electric dipoles (due to the spin-spin interaction). Comparing formulas (\ref{SDBEC disp for ED}) and (\ref{SDBEC disp for MD}) we see that they differ by the last term. Interaction of the electric dipoles leads to $d^{2}k_{z}^{2}$, where the external electric field is directed along the $z$ axes, but interaction of the magnetic moments gives $-\mu^{2}k_{\perp}^{2}$, where $k_{\perp}^{2}=k_{x}^{2}+k_{y}^{2}$ assuming that the external magnetic field directed along $z$ axes either.  This difference is caused by the fact that the electric and the magnetic fields satisfy to different pairs of the Maxwell equations.

For comparison of the last terms in the formulas (\ref{SDBEC disp for ED}) and (\ref{SDBEC disp for MD}) we show two functions $\xi_{d}=x\cos^{2}\theta$ and $\xi_{M}=-x\sin^{2}\theta$ on Fig.1 (a) and they modules of the Fig.1 (b).

\begin{figure}
\includegraphics[width=8cm,angle=0]{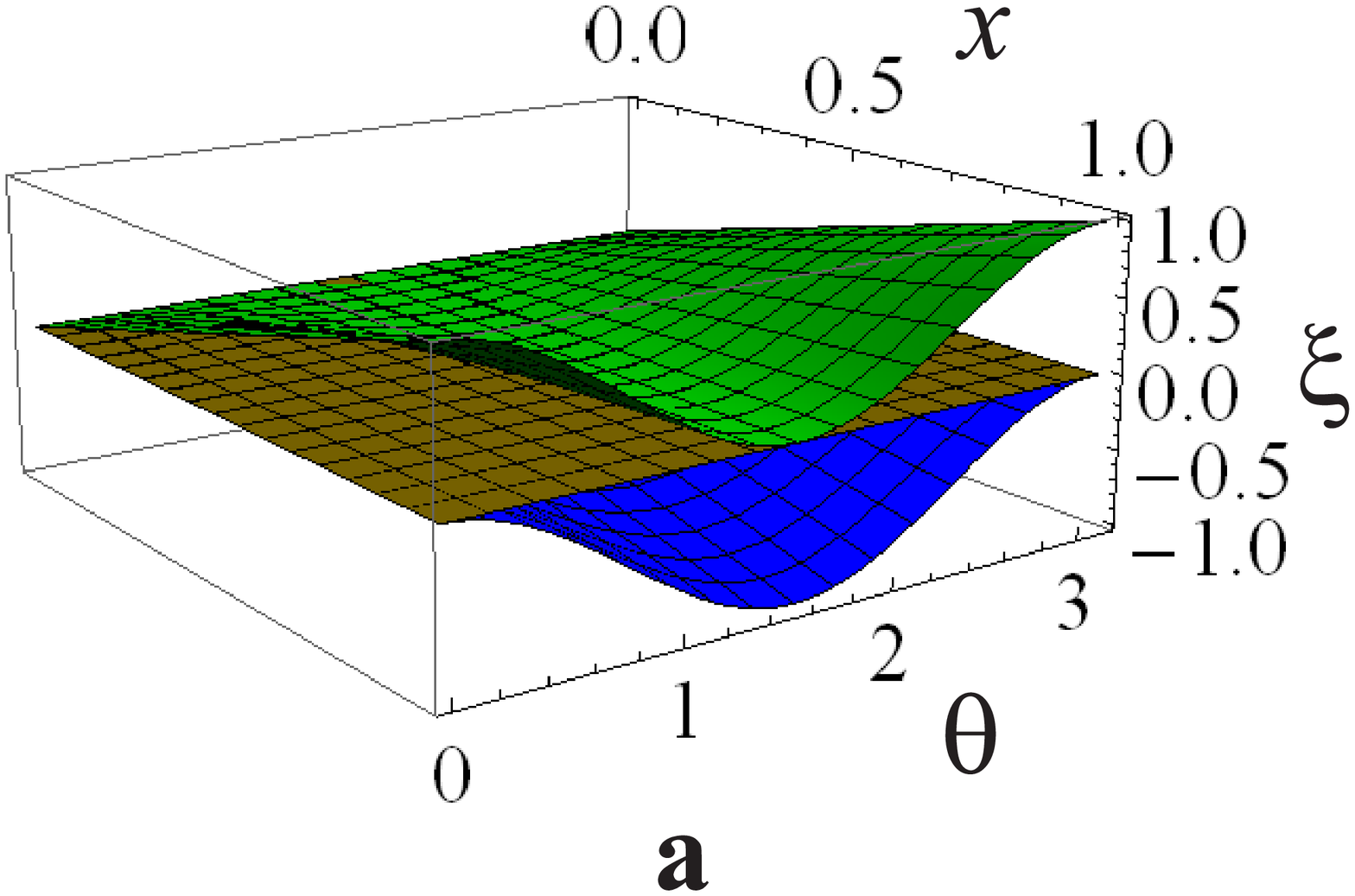}
\includegraphics[width=8cm,angle=0]{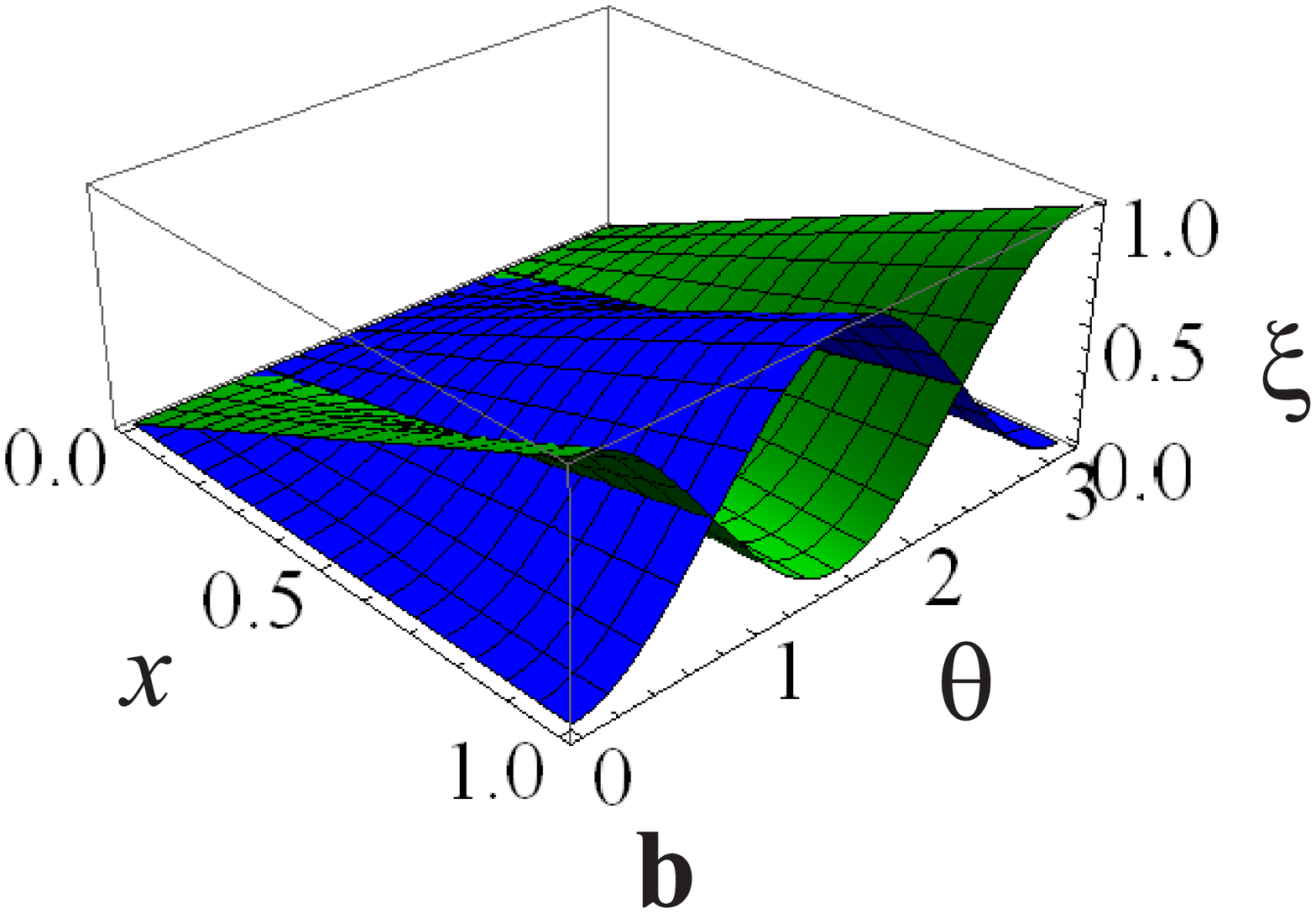}
\caption{\label{ESS 1} (Color online) The figure illustrates the dipolar part of
the dispersion dependence for the collective excitations in the dipolar BECs. We picture $\xi_{d}=x\cos^{2}\theta$ and $\xi_{M}=-x\sin^{2}\theta$. $\xi_{d}$ depicts the last term in formula (\ref{SDBEC disp for ED}) describing dispersion of the electrically polarized BEC. $\xi_{M}$ illustrates the last term in formula (\ref{SDBEC disp for MD}) describing dispersion in the magnetized BEC. Figure (a) shows $\xi_{d}$ and $\xi_{M}$. Plane on figure (a) presents unpolarized BEC. Figure (b) presents module of $\xi_{d}$ and $\xi_{M}$.}
\end{figure}

Formulas (\ref{SDBEC disp for ED}) and (\ref{SDBEC disp for MD}) differ from results have been obtained (see for instance \cite{Lahaye RPP 09} and \cite{Baranov CR 12}). There two results we have difference with. One of them obtained for the spinor BEC of spin-1 atoms \cite{Machida JPSJ 98}-\cite{Kawaguchi Ph Rep 12}. This difference is related with the fact that different constants of the short-range interaction were used in Refs. \cite{Machida JPSJ 98}, \cite{Ho PRL 98} for description of interaction between the species with different spin projection. We suppose that the short-range interaction is described with one interaction constant, but we include the long-range spin-spin interaction. Interaction of the magnetic moments with an external magnetic field is also introduced in some of these papers \cite{Szirmai PRA 12}-\cite{Zhang PRA 11}.
The magnetic field introduces two terms, one of which is given
by the linear Zeeman term and the quadratic
Zeeman term. These terms give the potential energy of particles with different projections of the spin being in the external magnetic field. So, we have considered different models for the spinor BEC description.

The second result was obtained on the way of generalization of the scalar GP equation, including in this equation a term describing dipole-dipole interaction \cite{Yi PRA 00}-\cite{Santos PRL 00}, \cite{Lahaye RPP 09}, \cite{Baranov CR 12}. It gives results similar to (\ref{SDBEC disp for ED}) and (\ref{SDBEC disp for MD}), but not the same. Comparison of formula (\ref{SDBEC disp for ED}) with results of other authors can be also found in Ref. \cite{Andreev 2013 non-int GP}. Let us discuss some reasons for this difference. Firstly, the generalized GP equation includes interaction of the magnetic dipoles and interaction of the electric dipoles in the same way, but we have shown that energies of the electric dipole interaction and the spin-spin interaction have different forms. This reveals in difference of the Maxwell equations for the magnetic field caused by the magnetic dipoles and the electric field caused by the electric dipoles. It also reveals in different properties of the collective excitation of the magnetized BEC and the electrically polarized BEC, as for linear excitations, their dispersion is presented by the formulas (\ref{SDBEC disp for ED}) and (\ref{SDBEC disp for MD}), and for the nonlinear soliton excitations described in the following sections.

\section{Solitons in unpolarized BEC}

Following Ref.s \cite{Fedele EPJ.B.02}, \cite{Andreev Izv.Vuzov. 10} we present analytical solutions for the bright, dark, and grey solitons in the unpolarized BEC, which will be generalized for the dipolar BECs in the next sections.

The bright and dark solitons have a trivial constant velocity field showing propagation of perturbation of the particle concentration. We have increasing of the particle concentration in the bright soliton existing in the attractive BEC, and decreasing of the particle concentration in the dark soliton existing in the repulsive BEC. The grey soliton exists in the repulsive BEC and reveals soliton structure as in the particle concentration and in the velocity field.

Presenting formulas appear as solutions of the following set of QHD equations for nonlinear plane waves propagating in the three dimensional unpolarized BEC, so equations emerge in the one dimensional form
\begin{equation}\label{SDBEC cont unpolar}\partial_{t}n+\partial_{x}(nv)=0,\end{equation}
and
\begin{equation}\label{SDBEC bal imp eq unpolar BEC}mn(\partial_{t}+v\partial_{x})v-\frac{\hbar^{2}}{4m}\partial^{3}_{x} n
+\frac{\hbar^{2}}{4m}\partial_{x}\biggl(\frac{(\partial_{x}n)^{2}}{n}\biggr)=-gn\partial_{x}n.\end{equation}

\subsection{Bright soliton}

The bright soliton solution appears for the attractive BEC. A three dimensional attractive BEC is unstable to the collapse. Therefore, for consideration of the attractive BEC we should consider the quasi-one-dimensional BEC. However, to get main picture of possible soliton solution, we can formally consider propagation of the plane soliton (one dimensional propagation) in the three dimensional BEC.

We suppose that the particle concentration
$n(x,t)$ depends on the space coordinate $x$ and time $t$ only in the combination
$\xi=x-V_{0}t$, where $V_{0}$ is the velocity of the bright soliton propagation. Assuming that perturbations caused by the soliton disappears at large distances from the soliton we come to the following boundary conditions
\begin{equation}\label{SDBEC Bound cond for bright}\lim_{x\rightarrow\pm\infty}n(\xi)=0.\end{equation}
We also assume that the velocity field is an arbitrary constant
$v(x,t)=V_{0}$.

One can obtain that the particle concentration in the bright soliton has form of
\begin{equation}\label{SDBEC bright soliton} n(\xi)=\frac{2m\mid E\mid}{\mid g\mid}\frac{1}{\cosh^{2}\biggl(\frac{\sqrt{2\mid E\mid}m\xi}{\hbar}\biggr)},\end{equation}
where soliton amplitude $n_{0}$ and phase $\alpha$ appear as
\begin{equation}\label{SDBEC bright soliton ampl} n_{0}=\frac{2m\mid E\mid}{\mid g\mid},\end{equation}
and
\begin{equation}\label{SDBEC bright soliton phase} \alpha=\frac{\sqrt{2\mid E\mid}m\xi}{\hbar},\end{equation}
correspondingly. In the formulas (\ref{SDBEC bright soliton})-(\ref{SDBEC bright soliton phase}) $E=c_{0}-V_{0}^{2}/2$ is an analog of the kinetic energy of the flow, but $E$ has minus in front of the velocity square, $c_{0}$ is a constant \cite{Fedele EPJ.B.02}, \cite{Andreev Izv.Vuzov. 10}.
$E<0$, $g<0$. Width of the soliton $D$ usually introduced as a distance between edges of the soliton at the half of its height. Hence, the width $D$ equals to $\hbar \ln(\sqrt{2}+1)/(m\sqrt{2\mid E\mid})$.

\subsection{Dark soliton}

The dark soliton exists in the unpolarized BEC with the repulsive short-range interaction $g>0$. It reveals as an area of rarefication of the particle concentration. We can come to this solution from the set of QHD equations (\ref{SDBEC cont unpolar}) and (\ref{SDBEC bal imp eq unpolar BEC}) or corresponding one-dimensional GP equation. For solution we have to assume that the particle concentration $n(x,t)$ depends on the coordinate $x$ and time $t$ in the following combination $\xi=x-V_{0}t$, hence $n(x,t)=n(\xi)$, where $V_{0}$ is a constant velocity showing velocity of the soliton propagation in the positive x-direction. Assuming that soliton gives no perturbation in the particle concentration at the infinite distance we write the following boundary conditions
\begin{equation}\label{SDBEC darc sol BC}\lim_{x\rightarrow\pm\infty}n(\xi)=n_{0}.\end{equation}
As for the bright soliton we assume that the velocity field is an arbitrary constant
$v(x,t)=V_{0}$.

The dark soliton solution appears as
\begin{equation}\label{SDBEC dark soliton} n(\xi)=n_{0}\tanh^{2}\biggl(\frac{\sqrt{mgn_{0}}}{\hbar}\xi\biggr).\end{equation}
We have deep rarefication in the center of the dark soliton since $\tanh(0)=0$. The particle concentration increases monotonically in both directions from the center of the soliton.

\subsection{Grey soliton}

The grey soliton in the unpolarized BEC, as the dark soliton, exists at the repulsive short-range interaction ($g>0$). It reveals as an area of rarefication of the particle concentration. In contrast with the dark soliton, we have a perturbation of the velocity field along with a perturbation of the particle concentration.

At consideration of the grey solitons we suppose that, as in previous cases, the particle concentration is a function of $\xi=x-u_{0}t$, so we have
$n(x,t)=n(\xi)$. We also assume that, there are perturbations of the velocity field $v(x,t)$ in vicinity of a velocity $V_{0}$, and this perturbation depends on $\xi$ either. Thus, we can write velocity field $v(x,t)$ as
$v(x,t)=V_{0}+V_{1}(\xi)$, where $V_{0}$, as in the previous subsection, is an arbitrary constant
current velocity associated with the Madelung's fluid
background motion and $V_{1}$ is an arbitrarily large current
velocity perturbation.

The particle concentration and the velocity field satisfy to the following boundary conditions
\begin{equation}\label{SDBEC grey sol BC conc}\lim_{x\rightarrow\pm\infty}n(\xi)=n_{0},\end{equation}
and
\begin{equation}\label{SDBEC grey sol BC vel}\lim_{x\rightarrow\pm\infty}V_{1}(\xi)=0.\end{equation}

It was shown that velocity of the grey soliton propagation should satisfy to the following conditions
$V_{0}-\sqrt{\frac{gn_{0}}{m}}\leq u_{0}\leq V_{0}+\sqrt{\frac{gn_{0}}{m}}$ (see Ref. \cite{Fedele EPJ.B.02}).

The particle concentration in the grey soliton appears as
\begin{equation}\label{SDBEC grey soliton concentration}n(\xi)=n_{0}  \Biggl(  1- A^{2} \frac{1}{ \cosh^{2} \biggl(\frac{\sqrt{mgn_{0} A^{2} }}{\hbar}\xi\biggr) } \Biggr),\end{equation}
where
\begin{equation}\label{SDBEC grey sol ampl}A^{2}=1-\frac{m}{gn_{0}}(u_{0}-V_{0})^{2},\end{equation}
and $0\leq A^{2}\leq1$. One can find
\begin{equation}\label{SDBEC grey soliton velocity}V_{1}(\xi)=\frac{(V_{0}-u_{0})A^{2}} {\cosh^{2}\biggl(\frac{\sqrt{mgn_{0}A^{2}}}{\hbar}\xi\biggr)-A^{2}}\end{equation}
for perturbation of the velocity field.

For the future references it is useful to introduce the phase of the grey soliton $\alpha_{g}$ as $\alpha_{g}=\sqrt{mgn_{0}}A\xi /\hbar$. Area of the maximum rarefication of the particle concentration is in the center of the grey soliton $n_{min}=n(0)$. If $A<1$ when $n(0)=n_{0}(1-A^{2})>0$. We see that the particle concentration does not necessarily equals to zero, whereas in the dark soliton $n(0)=0$. The particle concentration $n(\xi)$ monotonically increases in both directions from the center of the soliton. It is interesting to consider a limit case when $u_{0}=V_{0}$, then $V_{1}(\xi)=0$, $A^{2}=1$, and $n(\xi)$ (\ref{SDBEC grey soliton concentration}) simplifies to the dark soliton solution (\ref{SDBEC dark soliton}).
From figures (2)-(4) we see the form of the grey soliton. Figure (3) shows that the velocity field perturbation appears as an area of increasing (decreasing) of the velocity field at $V_{0}>u_{0}$ ($V_{0}<u_{0}$).

\section{Solitons in magnetized BEC}

This and next section are dedicated to solitons in the dipolar BEC. Magnetized BEC is described by the set of equations (\ref{SDBEC cont eq}), (\ref{SDBEC magn BEC Maxwell curl}), (\ref{SDBEC magn BEC Maxwell div}) and (\ref{SDBEC bal imp eq fully magn}). Electrically polarized BEC is governed by equations (\ref{SDBEC cont eq for ED}), (\ref{SDBEC field good curl}), (\ref{SDBEC bal imp eq short from GP}) and (\ref{SDBEC field good div simplified}). Solving the Maxwell equations for special cases we put these solutions in the Euler equation. Hence, the Euler equation represents in a simpler form, which is an analog of the Euler equation for the unpolarized BEC (\ref{SDBEC bal imp eq unpolar BEC}). Consequently, we can represent and use solutions, described above, for description of the dipolar BEC.

\subsection{propagation of the soliton perpendicular to external magnetic field}

Equation (\ref{SDBEC magn div short}) gives $B_{x}'=0$, where $'$ is the derivative on $x$, so we have $B_{x}=0$, as at infinity (in equilibrium) $x$ projection of magnetic field equals to zero. Equation (\ref{SDBEC magn curl short}) leads to two relations $B_{y}'=0$ and $B_{z}'=4\pi\mu n'$, the first of them gives $B_{y}=0$. Derivative of $B_{z}$ we put in the Euler equation (\ref{SDBEC bal imp eq fully magn}), and we find
\begin{equation}\label{SDBEC bal imp eq solMBEC per}mn(\partial_{t}+v\partial_{x})v-\frac{\hbar^{2}}{4m}\partial^{3}_{x} n
+\frac{\hbar^{2}}{4m}\partial_{x}\biggl(\frac{(\partial_{x}n)^{2}}{n}\biggr)=-\tilde{g}_{M}n\partial_{x}n,\end{equation}
where $\tilde{g}_{M}=g-4\pi\mu^{2}$. This equation corresponds to the spectrum of linear excitations obtained above. We see that contribution of the magnetization reveals as effective "attractive" changing of the short-range interaction constant.

\subsubsection{Bright soliton in magnetized BEC}

In the unpolarized BEC the bright soliton exists at the attractive short-range interaction. In considering limit the magnetization leads to additional attracting between particles, so magnetization of the attractive BEC does not break condition of the bright soliton existence. We have $g=-\mid g\mid$, consequently $\tilde{g}_{M}=-(\mid g\mid +4\pi\mu^{2})$. We see that for the system under consideration the magnetization leads to increasing of module of $\tilde{g}_{M}$. Only amplitude of the bright soliton depends on the interaction constant $n_{0}\sim 1/\mid \tilde{g}_{M}\mid$. Hence, we find decreasing of the bright soliton amplitude due to the magnetization of the attractive BEC.

\subsubsection{Dark soliton in magnetized BEC}

The dark soliton exists in the BEC with the repulsive short-range interaction ($g>0$), hence the magnetization leads to decreasing of the effective interaction constant down to $\tilde{g}_{m}=g-4\pi\mu^{2}$. First of all it reveals in the widening of the dark soliton. Using of the Feshbach resonance allows to change the of the interaction constant $g$. Thus, at small enough positive $g$. Contribution of magnetization can prevail. That leads to disappearing of the dark soliton, and on stability of the repulsive BEC to the collapse.

\subsubsection{Grey soliton in magnetized BEC}

Grey soliton solution in the unpolarized BEC presented by formulas (\ref{SDBEC grey soliton concentration})-(\ref{SDBEC grey soliton velocity}). In our case we have to replace $g$ with $\tilde{g}_{M}$. It changes the phase and "amplitude" $A$ of the grey soliton. To trace consequence of changing of the interaction constant we present Fig.s (2)-(4), where $n(\xi, g)$ and $V_{1}(\xi,g)$ are shown. From these figures we see that decreasing of the interaction constant due to magnetization leads to decreasing of the soliton amplitudes. We have the decreasing of the amplitude of the particle concentration and the amplitude of the velocity field. Figure (4) allows to compare profile of the particle concentration and the velocity field.

\subsection{propagation of the soliton parallel to external magnetic field}

Let us consider a nonlinear plane perturbation propagating parallel to external uniform magnetic field, which is directed along the $z$ axes. In this case space dependence of hydrodynamic variables $n$, $\textbf{v}$ and the magnetic field $\textbf{B}$ reduces to dependence on $z$. Equation $\nabla \textbf{B}=0$ (\ref{SDBEC magn div short}) represents as $B_{z}'=0$, where $'$ means derivative on $z$. We find that magnetic moments evolution gives no influence on perturbation propagating parallel to the external field in the magnetized BEC, since $B_{z}$ only gives a contribution in the set of QHD equations (\ref{SDBEC cont eq}) and (\ref{SDBEC bal imp eq fully magn}). Using equation (\ref{SDBEC magn curl short}) we obtain that $B_{x}$ and $B_{y}$ equal to zero. Finally, we have equation
\begin{equation}\label{SDBEC bal imp eq solMBEC par} mn(\partial_{t}+v\partial_{z})v-\frac{\hbar^{2}}{4m}\partial^{3}_{z} n
+\frac{\hbar^{2}}{4m}\partial_{z}\biggl(\frac{(\partial_{z}n)^{2}}{n}\biggr)=-gn\partial_{z}n,\end{equation}
which mathematically has no difference with equation (\ref{SDBEC bal imp eq unpolar BEC}). Consequently, solutions of equation (\ref{SDBEC bal imp eq unpolar BEC}) presented by formulas (\ref{SDBEC bright soliton}), (\ref{SDBEC dark soliton}), (\ref{SDBEC grey soliton concentration}) and (\ref{SDBEC grey soliton velocity}) are valid for our case and we can conclude that at solitons propagation along direction of the external magnetic field, the external field gives no influence on properties of solitons. They depend on the short-range interaction only. This conclusion corresponds to the properties of linear excitations in the magnetized BEC (\ref{SDBEC disp for MD}). From formula (\ref{SDBEC disp for MD}) we see that at propagation parallel to the magnetic field $\theta=0$ and the last term equals to zero. Thus, we have the unchanged Bogoliubov spectrum. So, nonlinear perturbation receive no influence of the magnetization at propagation parallel to the external field, as it was for the linear excitations (see section IV).

\begin{figure}
\includegraphics[width=8cm,angle=0]{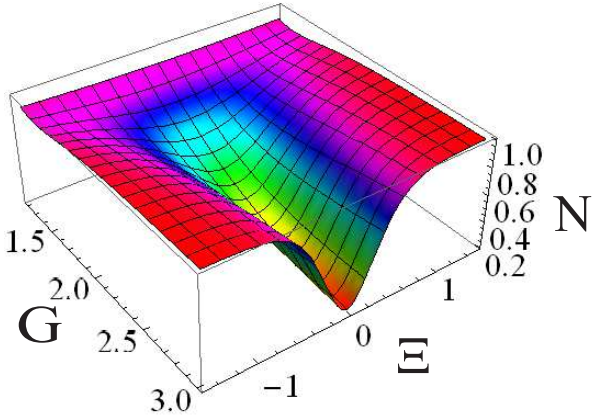}
\caption{\label{ESS 2} (Color online) The figure describes
the profile of the particle concentration $N=n(\xi)/n_{0}$ in the grey soliton. We consider the dependence of $N=n(\xi)/n_{0}$ on the reduced interaction constant $G=4n_{0}g/(mV_{0}^{2})$ and $\Xi=mV_{0}\xi/(2\hbar)$ at $u_{0}=V_{0}$.}
\end{figure}

\begin{figure}
\includegraphics[width=8cm,angle=0]{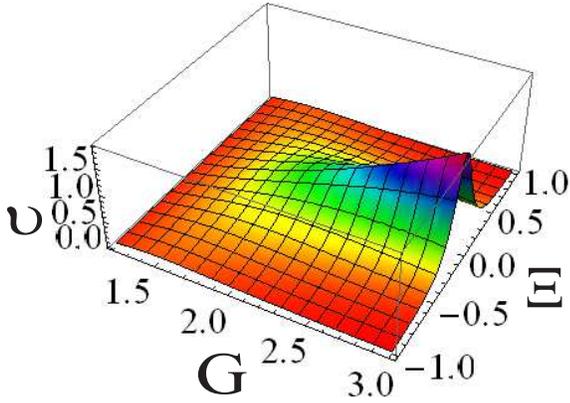}
\caption{\label{ESS 3} (Color online) The figure describes
the profile of the velocity field $\upsilon=V_{1}(\xi)/V_{0}$ in the grey soliton on the reduced interaction constant $G=4n_{0}g/(mV_{0}^{2})$ and $\Xi=mV_{0}\xi/(2\hbar)$ at $u_{0}=V_{0}$.}
\end{figure}

\begin{figure}
\includegraphics[width=8cm,angle=0]{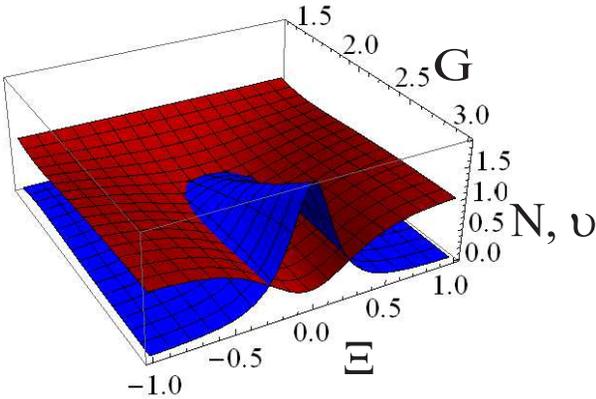}
\caption{\label{ESS 4} (Color online) On this figure we present
the profiles of the particle concentration and the velocity field on one figure to compare they structures. They are shown as functions of $G=4n_{0}g/(mV_{0}^{2})$ and $\Xi=mV_{0}\xi/(2\hbar)$ at $u_{0}=V_{0}$.}
\end{figure}

\section{Solitons in electrically polarized BEC}

We have described our results for the magnetized BEC. Let us present solitons in the fully polarized BEC of molecules or atoms having electric dipole moment. As we will see they have some different properties.

\subsection{propagation of the soliton perpendicular to external electric field}

Considering perturbation of hydrodynamic variables $n$, $\textbf{v}$, and $\textbf{E}$ are functions of $x-v_{0}t$ assuming that an external electric field is applied along $z$ axes. Therefore we consider propagation of the perturbation perpendicular to the external field. We find that the electric field gives no contribution in the evolution of the particle concentration $n$ and the velocity field $\textbf{v}$, the last term in equation (\ref{SDBEC bal imp eq short from GP}) equals to zero, since $\partial_{x}E_{z}=0$ that follows from equation (\ref{SDBEC field good div}), and we have no difference between considered here case and the unpolarized BEC. Consequently, we have that soliton solutions (\ref{SDBEC bright soliton}), (\ref{SDBEC dark soliton}), (\ref{SDBEC grey soliton concentration}) and (\ref{SDBEC grey soliton velocity}) propagate in the electrically polarized BEC perpendicular to the external field with no influence of the polarization.

%\begin{equation}\label{SDBEC bal imp eq solEPBEC per}mn(\partial_{t}+v\partial_{x})v-\frac{\hbar^{2}}{4m}\partial^{3}_{x} n
%+\frac{\hbar^{2}}{4m}\partial_{x}\biggl(\frac{(\partial_{x}n)^{2}}{n}\biggr)=-gn\partial_{x}n,\end{equation}

\subsection{propagation of the soliton parallel to external electric field}

We now consider propagation of a nonlinear perturbation parallel to the external electric field, so we suppose that $n$, $\textbf{v}$, and $\textbf{E}$ are functions of $\xi=z-v_{0}t$. In this case the last term in equation (\ref{SDBEC bal imp eq short from GP}) appears as $nd\partial_{z}E_{z}=-4\pi d^{2}n\partial_{z}n$ that we find using equation (\ref{SDBEC field good div}). Thus we can combine two terms in the right-hand side of equation (\ref{SDBEC bal imp eq short from GP}), which present $z$ projection of the force density $\textbf{F}$ and write it in the form of $F_{z}=-(g+4\pi d^{2})n\partial_{z}n$. We have that the dipole-dipole interaction of the fully polarized electric dipoles leads to change of the interaction constant. All difference, we find in compare with the unpolarized case, is replacement of $g$ with $\tilde{g}_{d}=g+4\pi d^{2}$. Final form of the Euler equation is
\begin{equation}\label{SDBEC bal imp eq solEPBEC par}mn(\partial_{t}+v\partial_{z})v-\frac{\hbar^{2}}{4m}\partial^{3}_{z} n
+\frac{\hbar^{2}}{4m}\partial_{z}\biggl(\frac{(\partial_{z}n)^{2}}{n}\biggr)=-\tilde{g}_{d}n\partial_{z}n,\end{equation}
Using methods described in Ref.s \cite{Fedele EPJ.B.02}, \cite{Andreev Izv.Vuzov. 10} we find changes of properties of the bright, dark, and grey solitons. General form of these solitons are given by formulas (\ref{SDBEC bright soliton}), (\ref{SDBEC dark soliton}), (\ref{SDBEC grey soliton concentration}) and (\ref{SDBEC grey soliton velocity}), but instead of $g$ we should put $\tilde{g}_{d}$ containing contribution of the electric dipole moment.

\subsubsection{Bright soliton in electrically polarized BEC}

In contrast with the magnetization, which leads to additional effective attraction $\tilde{g}_{M}=g-4\pi\mu^{2}$ at the wave propagation perpendicular to the external magnetic field, the electric polarization gives an effective repulsion $\tilde{g}_{d}=g+4\pi d^{2}$ revealing at wave propagation parallel to the external electric field. Consequently, polarization of the attractive BEC $g<0$ leads to decreasing of the module of the effective interaction constant $\tilde{g}_{d}=-\mid g\mid+4\pi d^{2}$. For the bright soliton it reveals in both an increasing of the soliton amplitude (\ref{SDBEC bright soliton ampl}) and a constriction of the area of soliton existence.

\subsubsection{Dark soliton in electrically polarized BEC}

The dark soliton can also exist in the electrically polarized BEC. At theoretical description it appears as a solution of equation (\ref{SDBEC bal imp eq solEPBEC par}) corresponding to formula (\ref{SDBEC dark soliton}). Increasing of the polarization leads to increasing of the effective interaction constant $\tilde{g}_{d}=g+4\pi d^{2}$. Consequently, the width of the dark soliton ($g>0$) $D=\hbar/(2\sqrt{m\tilde{g}_{d}n_{0}})\cdot\ln\frac{\sqrt{2}+1}{\sqrt{2}-1}$ decreases with the increasing of polarization. The amplitude $n_{0}$ of dark soliton is not affected by the interaction. It is determined by the equilibrium particle concentration.

\subsubsection{Grey soliton in electrically polarized BEC}

Let us describe changes of the grey soliton properties in the electrically polarized BEC in comparison with the unpolarized BEC. Fig.s (2) and (3) allow to get the consequence of polarization of the BEC on properties of the grey soliton. From figures (2) and (3) we see that the electric polarization of the BEC leads to increasing of the amplitudes of concentration and velocity field in the grey soliton. The width of the grey soliton increases as well. It happens due to increasing of the effective interaction constant $\tilde{g}_{d}=g+4\pi d^{2}$.

\section{Conclusions}

As a summary we recapture main results of our paper. We studied fundamental linear and nonlinear collective excitations both in the electrically polarized BEC and in the magnetized BEC. For this aim we have developed the QHD models for the fully polarized BEC, which are simplification of the QHD models for the dipolar BEC including evolution of the dipole directions we had developed in earlier papers. These models consist of the couple of QHD equations (the continuity and Euler equations). This couple of the QHD equations can be presented as a nonintegral nonlinear Schrodinger equation, which is the generalization of the well-known GP equation on dipolar BECs. The QHD equations or corresponding nonlinear Schrodinger equation have to be coupled with the Maxwell equations. The QHD equations for the electrically polarized BEC contain an electric field caused by electric dipoles of the medium, and, correspondingly, the QHD equations describing magnetized BEC contain an magnetic field caused by the magnetic moments. The electric field and the magnetic field satisfy to different pair of the Maxwell equations, consequently, we get differences in properties of the two kind of the dipolar BEC. These differences reveal in properties of the collective excitations obtained in the paper. Generalization of the Bogoliubov spectrum on the dipolar BECs contains a new term caused by dipolar effects. The electric polarization gives $k^{2}d^{2}\cos^{2}\theta$, whereas the magnetization leads  to $-k^{2}\mu^{2}\sin^{2}\theta$. These differ from well-known results obtained at using of the integral generalization of GP equation for the dipolar BEC. These differences are caused by mistreating of the potential energy of dipole-dipole interaction. Correct forms of the potential energy for the spin-spin interaction and interaction of electric dipoles are presented in the paper. We obtained that the generalization of the Bogoliubov spectrum are anisotropic. Form of anisotropy is similar for both kinds of dipolar BEC, but they were different. They have same form, but direction of axes of anisotropy differs on 90 degree. Anisotropic terms caused by dipolar effects have differen sign. Electric dipoles lead to effective repulsion, since we have $g+4\pi d^{2}\cos^{2}\theta$ instead of the interaction constant of the short-range interaction $g$. Magnetic dipoles give effective attraction, hence we obtained $g-4\pi\mu^{2}\sin^{2}\theta$ instead of $g$ for the unpolarized BEC. Our model allows to get several exact analytical soliton solutions for the dipolar BEC. We found solutions for the bright, dark, and grey solitons in the electrically polarized and in the magnetized BECs. We obtained solutions in cases when solitons propagate parallel or perpendicular to the external field. We discover that at propagation parallel (perpendicular) to the external magnetic (electric) field solitons in the magnetized (electrically polarized) BEC are not affected by the magnetization (electrical polarization), and they have same properties  as solitons in the unpolarized BEC. In opposite limits we calculated contribution of the dipolar effects in the amplitude and phase of solitons.

\section{Appendix}

We describe details of the calculation of the spectrum of collective excitations for the fully magnetized BEC.
From the continuity equation (\ref{SDBEC cont eq}) in the linear approximation, assuming that $n=n_{0}+\delta n$, $\textbf{v}=0+\delta \textbf{v}$, and $\textbf{B}=\textbf{B}_{0}+\delta \textbf{B}$, where $n_{0}$ and $\textbf{B}_{0}$ are equilibrium values of the particle concentration and the magnetic field, equilibrium value of the velocity field equals to zero, $\delta n$, $\delta \textbf{v}$, and $\delta \textbf{B}$ are small perturbations of the equilibrium state, for the Fourier amplitudes of the hydrodynamic variables, we find
\begin{equation}\label{SDBEC cont lin}\delta n=n_{0}\frac{\textbf{k}\delta \textbf{v}}{\omega}.\end{equation}
The Euler equation (\ref{SDBEC bal imp eq fully magn}) in the linear approximation represents as
\begin{equation}\label{SDBEC Euler magn lin}-\imath\omega mn_{0}\delta \textbf{v}+\imath \textbf{k} k^{2}\frac{\hbar^{2}}{4m}\delta n=-gn_{0}\imath \textbf{k} \delta n+\mu n_{0}\imath \textbf{k} \delta B_{z},\end{equation}
multiplying this equation on $\textbf{k}$ and using equation (\ref{SDBEC cont lin}) in the first term we get relation between the perturbation of particle concentration and z-projection of the magnetic field, which is
\begin{equation}\label{SDBEC Euler via concentration} -\omega^{2} m\delta n+ k^{4}\frac{\hbar^{2}}{4m}\delta n=-gn_{0} k^{2} \delta n+\mu n_{0} k^{2} \delta B_{z},\end{equation}
Equation (\ref{SDBEC magn curl short}) gives three scalar equations
\begin{equation}\label{SDBEC Max curl lin X} k_{y}\delta B_{z}-k_{z}\delta B_{y}=4\pi k_{y}\mu\delta n,\end{equation}
\begin{equation}\label{SDBEC Max curl lin Y} k_{x}\delta B_{z}-k_{z}\delta B_{x}=4\pi k_{x}\mu\delta n,\end{equation}
and
\begin{equation}\label{SDBEC Max curl lin Z} k_{x}\delta B_{y}-k_{y}\delta B_{x}=0.\end{equation}
From the equation (\ref{SDBEC Max curl lin Z}) we can get $\delta B_{x}$ via $\delta B_{y}$, using it and algebraic form of equation (\ref{SDBEC magn div short}), which is
\begin{equation}\label{SDBEC Max div lin} k_{x}\delta B_{x}+k_{y}\delta B_{y}+k_{z}\delta B_{z}=0,\end{equation}
we obtain $\delta B_{y}$ via $\delta B_{z}$ in the following form
\begin{equation}\label{SDBEC By} \delta B_{y}=-\frac{k_{y}k_{z}}{k_{\perp}^{2}}\delta B_{z}.\end{equation}
Next, we put (\ref{SDBEC By}) in equation (\ref{SDBEC Max curl lin X}) and find relation for $\delta B_{z}$, which is a part of the Euler equation (\ref{SDBEC Euler magn lin}), via the perturbation of the particle concentration as follows
\begin{equation}\label{SDBEC formula for Bz} \delta B_{z}=\frac{4\pi\mu k_{\perp}^{2}}{k^{2}}\delta n,\end{equation}
where $k_{\perp}^{2}=k_{x}^{2}+k_{y}^{2}$, and $k^{2}=k_{x}^{2}+k_{y}^{2}+k_{z}^{2}$. Therefore, we can write $k_{\perp}^{2}=k^{2}\sin^{2}\theta$, where $\theta$ is the angle between direction of wave propagation given by $\textbf{k}$ and direction of the external magnetic field $\textbf{B}_{0}$. Putting (\ref{SDBEC formula for Bz}) in the linearized Euler equation (\ref{SDBEC Euler via concentration}) we get formula (\ref{SDBEC disp for MD}).

\end{document}